\documentclass[sigconf]{acmart}

\makeatletter
\renewcommand\@formatdoi[1]{\ignorespaces}
\makeatother

\renewcommand\footnotetextcopyrightpermission[1]{}

\setcopyright{none}

\usepackage{comment}
\usepackage{endnotes}
\usepackage{xspace}
\usepackage{url} 
\usepackage{caption}
\usepackage{subfig}
\usepackage{color}
\usepackage{multirow}
\usepackage{enumitem}

\usepackage{float}
\usepackage{graphicx}
\usepackage{amsmath}
\usepackage[linesnumbered,ruled]{algorithm2e}
\SetKwRepeat{Do}{do}{while}%
\usepackage{algpseudocode}
\begin{document}

\title{Crashing Privacy: An Autopsy of a Web Browser's Leaked Crash Reports}

%
%

\setlength{\headsep}{6mm}

\author{Kiavash Satvat}
\affiliation{%
	\institution{University of Alabama at Birmingham}
}
\email{kiavash@uab.edu}

\author{Nitesh Saxena}
\affiliation{%
	\institution{University of Alabama at Birmingham}
}
\email{saxena@uab.edu}

	\begin{abstract}
	Harm to the privacy of users through data leakage
	is not an unknown issue, however, it has not been studied in the
	context of the crash reporting system. Automatic Crash Reporting
	Systems (ACRS) are used by applications to report information
	about the errors happening during a software failure.
	Although crash reports are valuable to diagnose
	errors, they may contain users' sensitive information.
	In this paper, we study such a privacy leakage vis-a-vis
	browsers' crash reporting systems. As a case study, we mine a
	dataset consisting of crash reports collected over the period of six years.
	Our analysis shows the presence of more than 20,000 sessions and
	token IDs, 600 passwords, 9,000 email addresses, an enormous
	amount of contact information, and other sensitive data. Our
	analysis sheds light on an important security and privacy issue
	in the current state-of-the-art browser crash reporting systems.
	Further, we propose a hotfix to enhance users' privacy and
	security in ACRS by removing sensitive data from the crash
	report prior to submit the report to the server. Our proposed
	hotfix can be easily integrated into the current implementation
	of ACRS and has no impact on the process of fixing bugs while
	maintaining the reports' readability.

\end{abstract}	

\maketitle
	
\section{Introduction}
\label{sec:intro}

A few studies were conducted, focusing on legal,
financial, reputational, and other aspects of information disclosure
\cite{romanosky2011data,acquisti2006there,cereola2011breach}. Some of these studies analyzed a specific data leakage incident \cite{acquisti2006there,cereola2011breach}.
Similarly, several studies analyzed the security and privacy of web browsers \cite{ter2008enhancing,satvat2014privacy,carlini2012evaluation,jackson2006protecting}. 
 However, no
previous study was conducted to examine the risks that the
current implementation of browsers' Automatic Crash Reporting System
(ACRS) may pose to the users' privacy.

A crash as an inevitable fact occurs when a software
terminates its normal execution process unexpectedly.
The use of ACRS is a common approach to collect crash reports from
the users. ACRS makes troubleshooting smooth amongst a
variety of factors contributing to the
crash, such as underlying hardware and associated software.
Using ACRS, companies collect crash reports
from clients to address application bugs and improve users'
experience during web browsing. This information is used by
developers to detect the crash root cause and rectify the bug.
  Chrome, Firefox, Safari, and Opera are a few common
names that use the ACRS to collect field crashes \cite{chrome,firefox,safari,opera}.
In web browsers, ACRS is designed to collect the stack trace
of the failing thread, user's system environmental information
(e.g., OS and browser version), and probable user's feedback
on the crash. 

However, a fact that has been widely disregarded is the
content of the crash collected from the users' systems and
the amount of private information that can be found in crash
reports which may pose a threat to the users' privacy.

 In this paper, to demonstrate possible risks that the current implementation
 of ACRS may pose to users' privacy,
we study a dataset of partially anonymized crash reports released by one of the major browsers\footnote{The name of the browser, year, and the fields of dataset are removed or anonymized as discussed in Section \ref{subsec:ethicalcons}}. The current state of the art browsers collect relatively similar data including browsed URL during the crash and runtime information, as they can be required for the debugging.  
To the best
of our knowledge, this is the first case study which
scrutinizes a database of 2.5 million partially anonymized crash reports, containing
visited URLs, system runtime information and users' descriptions. 
Conducting our inspection, we extracted
a significant amount of private data including, approximately
20,000 session and token IDs, 600 clear text password,
9,000 email address, and an enormous amount of contact
information, including names, addresses, and phone numbers.
The obtained results signify the deficiency of the
current approach and shows the potential for ACRS to
be a compelling target for attackers. Furthermore, to address this deficiency, we
proposed a hotfix to remove the sensitive information from the
environmental information, users' feedback, and visited URLs
at the client side prior to disseminating to the server.

\medskip
\noindent\textbf{Our Contribution.} 
The detail contributions of our system are as follow.

\begin{itemize} [leftmargin=*]
	\item \textbf{Data Analysis (Section \ref{sec:dataanalysis}).} We conduct an empirical study
	on partially anonymized crash reports released by one of the
	major web browsers. Through our inspection, we extract
	and represent a significant amount of personal information
	and confidential data presented in the reports, delineating
	to what extent the current implementation of ACRS can be
	harmful to users' security and privacy. Our analysis sheds light on an
	important security and privacy issue in the current state-of-the-art browser crash reporting systems. 
	
	\item \textbf{Deploying A Hotfix (Section \ref{sec:countermeasures}).}  According to the result
	of our data analysis, we propose a hotfix, to enhance users'
	privacy and security in ACRS, by removing sensitive data
	from the users' description field and URL. Our hotfix can be
	easily integrated into the current implementation of ACRS
	and has no impact on the process of fixing bugs while
	maintaining the reports' readability. However, it protects
	users' private data against unauthorized access in the cases
	server gets compromised or data get leaked.
\end{itemize}
  
\noindent\textbf{Paper Organization.} The rest of this paper is organized as follows. Section
\ref{sec:background} provides the background about ACRS and explains the dataset used in this study. Section \ref{sec:methodology} defines the methodology used to analyze the data and presents the reports. In Section \ref{sec:dataanalysis}, we present the results of our data analysis. Section \ref{sec:countermeasures} describes insights and countermeasures. Finally, Section \ref{sec:conclusion} concludes our study and
suggests future research directions.

	\vspace*{2mm}
\section{STUDY PRELIMINARIES AND DATASET}
\label{sec:background}

\subsection{ACRS Background}
The use of ACRS is a common approach for collecting and
handling crashes. ACRS primarily consists of two main components.
First, what we refer to as Collector, a set of client side
libraries which collects the data from the clients and transfers
it to the server for further processing. Second, the Processor
which is a set of servers and services responsible for analyzing,
categorizing, and reporting the collected crashes. On the
client side, Collector, as an interacting interface between the
Processor and clients, runs when the application terminates
its normal execution process. Collector typically asks for the
users' permission to send the crash report to Processor or
to quit without sending the report. While choosing to send
the report to the Processor, the user has an option to include
the visited website at the time crash occurred or not. The
user also can provide additional information in the description
field to explain the issue that caused the crash. Additionally,
there might be an option for the user to provide an email
address for future support and contributions. After these steps,
Collector transfers a dump of crash report and collected details
to Processor that is responsible for further processing and
presenting for the view of developers and the general public.  

 The collected data from clients are referred to as Crash reports. Each crash report is composed of two
components. The first component is the stack trace of the
failed thread which is generally referred to the minidump and
carries information about the system memory. The second
component is the runtime information of the users' system
and the possible feedback provided by the user. While these
details are essential for the developers to detect where the
problem lies, they may contain users' sensitive information.
For instance, the minidump as a sensitive chunk of data
may contain information such as username, password, and
encryption key \cite{broadwell2003scrash}. Also, it is not hard to expect that the
runtime information and users' comment may carry private
information such as contact details and other sensitive data.

\subsection{Dataset} \label{subsec:dataset}
In this paper, we study a  dataset
of 2,493,278 partially anonymized crash reports consists of
details such as visited URLs, time of the crash, the client operating
system, and user description of the crash. A dataset aggregated by a major web browser during the course of 6 years. The current approach employed by the software companies, including web browsers (e.g., \cite{Firefoxsimilar,Chromesimilar}),
stores similar data in crashes.
Our close inspection revealed that while some of the fields
were fully anonymized (e.g., field \texttt{email}), unsuccessful
de-identifications of the other fields left a notable
amount of sensitive and personal information in the database. Private
data like the usernames, passwords, and emails embedded
in the URL field, or some confidential information in the
description field which is shared by the users.
We noticed the presence of three types of data in this dataset;

\begin{itemize}[leftmargin=*]
	\item \textbf{Deleted Fields:} Fields like \texttt{email} and \texttt{login} were deleted since they explicitly carry sensitive information.  
	
	\item \textbf{Masked Fields:} In this category, some efforts have been taken for data de-identification. In  \texttt{ip}, some records were de-identified by replacing the last two digits of IP address with zeros. 
	
	\item \textbf{Untouched Fields:} All fields except \texttt{email}, \texttt{login} and \texttt{address} apparently remained  untouched, however, not necessarily all of them convey any meaningful information. In this paper, we mostly focused on these attributes to demonstrate the breach of the information. 
	
\end{itemize}

\section{Methodology}
\label{sec:methodology}

In this section, we explain our approach for inspecting and analyzing the dataset. We also describe the methodology we use to present our results in Section \ref{sec:dataanalysis}.

\subsection{Ethical Consideration} \label{subsec:ethicalcons}

We conducted our research based on a dataset which
	was initially published and later removed following a report
	regarding the presence of sensitive information. During our
	scrutiny, however, we noticed the presence of similar data
	which is widely available for the public use. Therefore our
study aims to highlight a potential pitfall associated with
the current implementation of the ACRS to the researchers,
software developers, and the society.
Our study neither propagates the data nor does it draw
attention to any specific company or individual. Hence, we
believe our study causes no further harm to victims. To
preserve the privacy of associated individuals, companies and
third parties, and avoid working directly with this sensitive
information, we used complex queries and avoided performing
a manual check. Therefore, all results provided in Section
 \ref{sec:dataanalysis} represents a close estimation based on the output of our
queries. For the presentation, we removed identifiable information
from our results, including but not limited to the names
of individuals, websites, applications, companies and their
contact information, such as phone numbers and addresses.
The de-identification process was performed using a python
script replacing this sensitive information with asterisks.

Similarly, we removed all dataset related details which may imply a vendor or result in identifying an individual including  names, years, and other traceable information. We also changed the dataset's fields' names to avoid defaming company by possible cross referencing of presented data in this paper.

\subsection{Our Approach} \label{subsec:ourapproach}

\textbf{Main Fields of the Database.} 
In this study, we mainly focus on two fields of \texttt{url} and \texttt{description} which carry more sensitive information. 

\begin{itemize} [leftmargin=*]
	\item \texttt{\textbf{description.}} To understand what might be laid in this field we considered the following questions: \textit{``What are the possible users' feedbacks on crashes?''} ,\textit{``What might be the response of an expert user to the crash?''}, \textit{``Is it possible that a user shares his private information over the description field?''}.  According to these questions, we mined the fields, using keywords such as ``username'', ``password'', ``my phone number'', and ``please contact me'' to inspect the presence of gripping information
	and the data that may impact users' privacy.
	
	\item \texttt{\textbf{url.}} Similarly, for the URL field, we considered the following questions: \textit{`` what sort of information the URL can carry?''},   \textit{``Is it possible to find any sensitive information embedded in URL?''},  \textit{``What if the browser crashes at the time that user presses the login button to access his account?''},   \textit{``What if the browser crashes when a user presses the submit button to register to a website or upload a form?''},   \textit{``What if the browser crashes when an attacker was launching an attack against a website?''}.  According to these  questions, and using keywords
	such as ``username='', ``password='', we mined this field to extract those records that may be harmful to users' security and privacy.     
	
\end{itemize}

\noindent \textbf{Cross Referencing.} We performed no cross referencing between
the different fields of the database or with other auxiliary
resources such as social networks or previously published
datasets. We believe that doing so can result in exposure of
more private information. For instance, an attacker can locate an unpopular/distinct platform  (platforms that are being used in a particular industry e.g., banks) by cross referencing between the
database fields, such as \texttt{ip}, \texttt{platform},
\texttt{language}. Using these fields, the attacker can
obtain the approximate location (e.g. city), and then spot the
exact location using the date and platform fields.

\subsection{Categorizing Data Breach Severity} \label{subsec:categorizingDBSeverity}

A variety of risk analysis methodologies are proposed to
appraise and calculate the breach severity and the impact
of the associated risks \cite{stiennon2013categorizing,enisa}. 
These methods consider 
several factors, including context or type of the data, source
of the leakage, and potential impact to rank the severity of
each incident. These methods are applicable to all previously
reported incidents, as the majority of these leakages consist of one type of entity. For instance, in the case of the JPMorgan \cite{JPMorgan}, as one of the most severe data breaches of all time \cite{topbreaches},
 users' contact information breached. In the case of Myspace data breach \cite{myspace}, subscribers' account details were
subjected to the breach. This similarity makes the assessment
and evaluation of the associated risk relatively easy. However,
in the case of browsers data leakage, we deal with a set of
dissimilar and independent records, each being generated by a
different user and in a disparate circumstance. 
For instance,
while a record may only carry simple data about user's
browsing, the other may convey user's medical condition
or financial details.
 This diversity adds complications and
gives each of these records its own unique characteristic and
makes the prior methodologies defective towards this dataset.
Therefore, to present the result of our inspection, we define
three categories of risk severity that are designed to address our
specific needs and are compatible with the current dataset. We
present the data based on their severity into three categories:

\begin{itemize} [leftmargin=*]

	\item \textbf{Low Risk:} This category presents those records which does not
	impact individuals' privacy, but accessing them may provide
	additional information to an attacker, helping to launch
	targeted attacks against specific population (e.g., users who
	are using a particular operating system in a small region
	or a specific range of IP address). This information may
	also be interesting to the researchers and developers as they
	statistically represent the web population.
 	\item \textbf{Medium Risk:} This category presents those records which do
 not compromise users' privacy or security, but
 a cross referencing with an auxiliary information can turn them into a threat to the users' privacy.

 \item\textbf{High Risk:} This category presents records which explicitly violate users' privacy and contains users' sensitive information such as username, password, and contact details. Similar to the data presented in the previous categories data appeared in this category can be shared by users through description or URL field.

\end{itemize}

%

\section{Data Analysis} \label{sec:dataanalysis}

In this section, we demonstrate the results of our data analysis and present them based on the severity level introduced in Section \ref{sec:methodology}. 

\subsection{Low Risk} \label{subsec:lrisk} 

As discussed in Section \ref{subsec:categorizingDBSeverity}, under the low risk category, we present information which does not violate privacy, but at the same time is highly descriptive in terms of providing statistical details, and can be a representative of the whole Internet users population.
In contrast to reports offered by measurement websites (e.g., \cite{w3statistic,netmarketshare}), our dataset encompasses a wide range of users, while they were visiting various websites. Moreover, the data has been collected over the course of six years and has no correlation with any specific application, company, or individual, which signifies the impartiality of our dataset.

\noindent\textbf{Platforms and Protocols.} 
Market share websites (e.g., \cite{osmarketshare,w3osshare}) are the main
source, to obtain statistic on the popularity of different platforms
on the Internet. These websites build their database using
the obtained data from their visitors' browsers. Unlike these
reports, the random distribution of users in the current dataset
provides a more accurate statistic on the popularity of different
platforms. In this study, such information were extracted from the \texttt{platform} field. Similarly, same approach can be taken to extract the of various web protocols amongst websites.  Unlike other reports, which target the specific domains (e.g. most popular websites \cite{semrush} ) , the current dataset formed from a wide range of users and can be counted as a small sample of the whole web population.

\noindent\textbf{Identified Attacks.}
Considering the fact that a significant number of attacks against websites are launched through URL, we queried the \texttt{url} to detect the presence of
common attacks such as SQL Injection, Local File Inclusion,
and Directory Traversal. Below are some of the known attacks
we found during our inspection.

\begin{itemize} [leftmargin=*]
	\item \textbf{SQL Injection:} We used a combination of keywords such as ``select'', ``from'', ``where'' and ``union'' to detect the presence of SQL injection. 
	Below URL illustrates a situation in which the attacker's browser crashed while he was launching a SQL attack against a website.

\medskip

\texttt{http://www.******.com/photo/gallery/search.php?\newline search\_user=x\%2527\%20union\%20select\%20user\_password\newline
	\%20froum\%204images\_user\%20where\%20Andrew\%20Whyte=\newline
	\%2572}

\smallskip

	\item \textbf{Directory Traversal:} In this form of attack, the attacker can get access to restricted directories of the web servers (e.g., root directory). The attacker can use expression ``../'' to instruct the system to move between directories. To look up this type of attack, we searched for ``../'' as a keyword. 
	The following is an instance of this attack found in our database.  

\smallskip
{
	\mdseries\itshape\urlstyle{tt}
	\url{http://www.*****.com/de_old/index.html?prm_popup=../../}}
\smallskip

	\item \textbf{Phishing Attack:} Since it was not possible to search for all the websites, except by querying the \texttt{url}, we tried to inspect the users feedback on the crash. The below example is a paradigm where the user was trying to report a phishing website. 

\smallskip
{
	\mdseries\itshape\urlstyle{tt}
	\url{http://******/www.wellsfargo.com/
}}
\smallskip

\end{itemize}

Similarly, we used a same approach to extract other suspicious records. For instance, we searched for ``/etc/passwd/'' as a keyword for detecting the Local File Inclusion or ``document.cookie'' for the Cross Site Scripting (XSS).

\subsection{Medium Risk} \label{subsec:mrisk} 

\noindent\textbf{IP Address.} We observed three types of IPs in \texttt{ip}. The first type contains IPs which were fully deleted, the second type consists of records which were masked by turning into ``10.2.0.0'' and finally 261,388 records that were partially anonymized by replacing the last two blocks of IP with zeros. Moreover, a close inspection of the \texttt{url} revealed a significant number of URLs with embedded IP addresses. The below sample shows an IP address inside a URL.

\smallskip

\texttt{http://******/?getpostdata=get\&name=******\&site=\newline
	submit\&ip=**.**.**.**\&ref=000/}

\smallskip

\noindent\textbf{Email.} We extracted 9139 emails including personal and business email addresses. Out of which 7731 email obtained from the description field, while the remaining 1462 extracted from the URLs. The below sample shows an embedded email inside the URL.

\smallskip
\texttt{https://*****.net/activate.php?user=*****137199\&\newline
	email=*****@gmail.com/}

\smallskip

\noindent\textbf{Location.}
A fraction of users could have been located based on the data in \texttt{url} and \texttt{description} fields. 
locating users can occur either through the URL (e.g. when the user was trying to submit a registration form with an embedded address in URL) or by querying the \texttt{description} for the cases which user shared his address.

\subsection{High Risk} \label{subsec:hrisk}

\noindent\textbf{Username and Password.} We were able to pinpoint over 621 passwords. This number of users' credentials further escalate the severity of this leakage when we consider the fact of the pervasiveness of password reuse (43-51\%) \cite{das2014tangled}.
Some of these passwords were extracted from the URL, and the rest were embedded in the description field. In the case of URL, the credentials were exposed due to the weaknesses in design and development of the application where both the HTTP and GET method have been used to transfer the sensitive data between client and server while this method of transferring has been refused by the standards \cite{getrequest}. Below URL demonstrates a case where the browser crashed when a user was about to log into a website.

\medskip
\texttt{http://******/logon.do;jsessionid=abbTXFV3-1BSw7?\newline
		username=******\&password=******/}

\medskip

The next series of passwords were extracted from the \texttt{description}, where the user shared his password in the description field of Crash Reporter aiming to receive future support.

\medskip
\texttt{I can login but the ****** don't remember password for this site.
	I tell you the password to login:
	username: ******
	password: ******
	you can login south west of page. if you need translation go to http://www.*****.net/dic/.}
\medskip

\noindent\textbf{Token \& SessionID.} Stolen SessionID within its life cycle,
means that the account is stolen \cite{wu2015web}.  Our inspection revealed 21298 session and token IDs. Though at the time of writing this paper we cannot verify the persistence of vulnerability, considering the fact that sessions authenticate users, we can understand the risk posed to individuals' security and privacy at the time of the crash. Moreover, in our analysis, we noticed the presence of 7000 Tokens (as more severe authentication method compared to session IDs). The below URL depicts one of the samples with token Id.

\smallskip

\texttt{
https://******.com/*****/Main/Login\_WS.aspx?\newline
tokenid=880217f4-94c3-496d-8628-b2388b4ef299}
\smallskip
   
\noindent\textbf{Contact Information.} 
During our data analysis, we dealt with a significant number of
contact information some laid in URL while the others
were provided by users intentionally in the description field
to receive further support. Among these records, the latter
is specifically more threatening as can be used to launch a
targeted attack against the inexpert user. The attacker with access
to such records may easily undermine users' security and
privacy by contacting the user as a member of browser's
support team to deceive the user to install a Malware 
or request additional information from the target

\smallskip
\texttt{Please call me at ****** to discuss
	details about missing info on my ebay bid
	pages. Never a problem in the past, now an
	everyday problem.}
\medskip

Similarly, we noticed a considerable number of  phone numbers laid in \texttt{url}. Below demonstrates a number embedded inside  URL.

\medskip

\texttt{http://******/jcp/OrderDetail.aspx?context=OrderHist\newline
	\&iia=ya3z*****ZTtG\&pid=1\&OrderNum=******\&Phone=******/}

\smallskip

\section{Insight and Countermeasures } \label{sec:countermeasures}

A variety of direct and indirect factors contributed to the
privacy harm caused by this data leakage incident. Factors such
as disregarding the importance of data retention and its impacts
on users' privacy and security \cite{blanchette2002data,crump2003data,bignami2007privacy}, improper use of
GET method and the human error \cite{liginlal2009significant,wood1993human,whitman2003enemy} in disclosing
users' information. While these kinds of factors are widely
discussed, yet they are inevitable and as the result, we witness
an unprecedented wave of data leakages. Therefore, in this
section, we propose a hotfix that can be deployed into the
current ACRS, helping to enhance the security and privacy
of users against such a data disclosure, by removing sensitive
information from the crash report.

\subsection{Client Side Data Sanitization} \label{clientdatasanitizatio}

In the case of ACRS, both server and client side approaches
can be employed to enhance users' privacy and security
by removing sensitive information from the reports. While
removing the users' sensitive data at the server side
may provide more flexibility for the developers to get the
best readability of a report, the user should not be obligated
to trust the developer and the system to send their sensitive
data to the server. Therefore, in this paper, we propose a
client side approach to remove sensitive data from the crash
report's environmental information, aiming to safeguard users'
sensitive data right at the client side. This approach, as a
hotfix to the current ACRS implementation, can be easily
integrated with the popular ACRS distributions (\cite{Breakpad,Chrommium}),
aiming to sanitize users' private information such as username,
password, and social security number without any impact on
the bug fixation and with minimal or no impact on the 
readability of reports.
Unlike \cite{broadwell2003scrash} and \cite{castro2008better} which only focused on sanitizing minidump, we consider the crash runtime information and supportive data including description field and URL.
Similar to \cite{broadwell2003scrash}, we hypothesize that developers do
not want to get occupied by users' sensitive data.

To sanitize a report at the client, we developed a light weight
script which parses the URL and user's description field and
removes the private information prior to transferring them to
the Processor. The program masks the possible sensitive data
from the crash report based on the predefined list of sensitive
data (which can be defined by developers). The program takes
URL and description as input and checks them against
an array of predefined sensitive keywords which may appear
in those fields. In case of any match, the program masks the
following 4 characters of the sensitive keyword. The result
will be a sanitized report where sensitive data is unusable (or
difficult to guess) in case of an unauthorized access, but at the
same time the report maintains its readability for the debugging purpose. 

We define readability as a situation where developers are able to replicate the crash. For instance, in case of a browser crash, developers may require the visited URL in addition to stack trace information. From the URL developers are specifically looking for details like the name of the visited website and status where the crash occurred (e.g., posting a form). While our hotfix leaves these details untouched, it masks users' sensitive information which may exist inside URL and is not in the interest of developers (similar to \cite{broadwell2003scrash} hypothesis). 
Similarly, using our hotfix, a URL that carries token ID
will be unusable by attacker for the purpose of login, as its first
4 digits are masked, but it still denotes the necessary details
which are required for the debugging purpose.
 Depending on
the URL or description field, the list may include keywords
such as username, password, token, and session ID. The
list of keywords can be defined by developers, giving them
the opportunity to make a balance between users' privacy
and readability required for the crash debugging. Algorithm
1, represents our proposed method for removing private data
from the report and description field.

As a preliminary evaluation, we tested our program, against
500 independent records from \texttt{description} and
\texttt{url}. To preserve users' privacy during the testing
phase and to make sure that the data is not traceable back to
a user, we take some steps as follow. First, we tested each
field independently to ensure that data is segregated from the
other fields in the same record and therefore is not traceable
back to users. Second, for the case of URL sanitization, we
replaced the domain names with the \texttt{www.example.com}
(replacing the domain names with the \texttt{www.example.com}
was only performed for the testing purpose. The application
in the normal process keeps the domain untouched as the
developers may need it for the further analysis). After running
our program against de-identified records, we performed a
manual check to  assure that the URLs' readability is not
compromised. The result of our manual check showed that the
program was fully successful in manipulating the keywords.

The results of our evaluation on 500 URLs carrying session ID and 500 description fields carrying password show
100\% success rate on masking sensitive information from 
both fields. If we define readability such that the program
still maintains all information required to determine a crash
root cause from the URL, including the name of the website and
possible application or directories, the readability rate in the
case of URL was over 92\%. 
The result of our manual check
indicates that readability of 40 URLs out of 500 URLs has been impacted. As in some cases URL sanitization removed insensitive characters from the URL or caused removal of the accessed directory on the web application.

However, in the case of the description field, the result cannot
be quantitatively evaluated due to the qualitative nature of
the readability factor of the description field. To estimate the
readability rate of the description field a study should be
conducted to ask developers with authorized access to the
original and masked data to compare and rank the readability
of the descriptions. Given that such a study may jeopardize the
privacy of the users, we were not able to conduct this study.

\begin{algorithm}[!t]
	\KwData{$input$ (url, description)}
	\KwResult{anonymized \textit{input}}
	initialization;\\
	lst = [x,y,z]; \\
	open input;\\
	\For{\texttt{<line in input>}}
	{
		\For{\texttt{<i in lst>}}
		{
			regex = ("lst[i]\textbackslash \textbackslash W+(\textbackslash \textbackslash w+)");\\
			match= findall(regex, line);\\
			line = line.replace(match, "****");\\
		}
		
	}
	\caption{Sanitizing Crash Report}
	\label{alg:algorithm}

\end{algorithm}

The proposed approach soundly works in the case of
sanitizing URL, since the cases from which the keywords
are derived are relatively limited. However, the mass filtering
of data may result in removing insensitive information from
the description fields. As an example, developers choices for
defining sessionID in URL are likely limited to \texttt{sessionid=},
\texttt{session=}, and \texttt{sid=}. But in the case of users' description,
we deal with a great volume of uncertainty due to the infinite
ways of writing and explaining a topic. Spelling, use of
abbreviation may also add more complication to the situation.
For instance, while description \texttt{``I saved my username
in browser, but using the saved password
it doesn't get through the website! looks
the password is wrong!''} contains no sensitive
data, mass filtering turns it to \texttt{``I saved my username****rowser, but using the saved password
****oesn't get through the website! looks
saved password ****rong!''}. It is not hard to guess
that a considerable fraction of the descriptions may follow
the same pattern and the outright filtering can result in the removal
of insensitive data. We were also unable to provide a report
on the frequency of such cases, since the process requires
validation and manual comparison of the program's output
with the original report which would raise ethical concerns
and may violate users' privacy.

Moreover, it should be considered that the impact on the 
readability of the insensitive data remains negligible towards the
final process of fixing bugs considering the huge number of
crash reports received by the web browsers. In fact, many
of the submitted reports (approximately 90\%) are tagged as
duplicates, and are not being processed \cite{Socorroblog,ahmed2014impact}. Further
studies are required to examine the result of applying more
advanced data sanitization techniques (e.g., \cite{mivule2013comparative,li2015iterative,amiri2007dare,oliveira2003protecting}) on the ACRS, including the use of natural language
processing and deep learning. However, compared to our
approach, applying these techniques may result in a total loss
of a report when it is classified as sensitive and would be computationally expensive which may affect the system performance.

%
	
\section{Conclusion}
\label{sec:conclusion}

The plethora of data leakages has received a close review
by the media and research communities. However, no previous
study has been conducted in the context of browsers and
the crash reporting systems. In this paper, we studied the
disclosure of 2.5 million records of browser crashes that had
been collected by one of the major browsers. In our inspection,
we showed the potential privacy and security harm the current
implementation of ACRS may pose. Our work presents a
crucial preemptive step to raise community's awareness of
the security and privacy risks associated with the current
implementation of ACRS.
Further, we proposed a hotfix which masks the users' sensitive
data presented in the URL and description field without
impacting the readability of the reports. Our proposed hotfix
can be integrated with the current implementation of
ACRS to safeguard users' private data. While our proposed
approach finely works toward protecting users private information,
further studies are required to examine more advanced and
intelligent techniques, including natural language processing
and deep learning techniques.

%
		{\footnotesize \bibliographystyle{ACM-Reference-Format}
			\bibliography{bibliography}}

%
	
\end{document}